\documentclass[12pt]{article}
\usepackage{graphicx}
\usepackage{dcolumn}
\usepackage{amsmath}
\usepackage{units}
\usepackage{pstricks}
\usepackage{multirow}

\newcommand{\optbar}[1]{\shortstack{{\tiny (\rule[.4ex]{1em}{.1mm})}
  \\ [-.7ex] $#1$}}
\def\BorBbar    {\kern 0.18em\optbar{\kern -0.18em B}{}\xspace}
\def\DorDbar    {\kern 0.18em\optbar{\kern -0.18em D}{}\xspace}
\def\KorKbar    {\kern 0.18em\optbar{\kern -0.18em K}{}\xspace}

\newcommand{\BABARPubYear}    {04}

\newcommand{\BABARConfNumber} {044}
\newcommand{\SLACPubNumber} {10644}

\input{pubboard/babarsym}

\long\def\inst#1{\par\nobreak\kern 4pt\nobreak
    {\it #1}\par\vskip 10pt plus 3pt minus 3pt}

\begin{document}
{\pagestyle{empty}

\begin{flushright}
\babar-CONF-\BABARPubYear/\BABARConfNumber\\ 
SLAC-PUB-\SLACPubNumber\\
\end{flushright}

\par\vskip 2.0cm

\begin{center}
\Large \bf \boldmath
Measurements of Branching Fractions and \CP-Violating Asymmetries in $B$-Meson
Decays to the Charmless Two-Body States $\Kz\pip$, $\Kzb\Kp$, and $\KzKzb$
\end{center}
\bigskip

\begin{center}
\large The \babar\ Collaboration\\
\mbox{ }\\
August 17, 2004
\end{center}
\bigskip \bigskip

\begin{abstract}
We present preliminary measurements of branching fractions and \CP-violating 
asymmetries in decays of $B$ mesons to two-body final states containing a \Kz.
The results are based on a data sample of approximately $227$ million
\upsbb\ decays collected with the \babar\ detector
at the \pep2\ asymmetric-energy $B$ Factory at SLAC. We measure 
$\BR(\Bp\to\Kz\pip) = (26.0 \pm 1.3 \pm 1.0)\times 10^{-6}$,
$\BR(\Bp\to\Kzb\Kp) = (1.45 ^{+0.53}_{-0.46} \pm 0.11)\times 10^{-6} (< 2.35\times 10^{-6})$, and
$\BR(\Bz\to\KzKzb) = (1.19^{+0.40}_{-0.35}\pm 0.13)\times 10^{-6}$, where the
first uncertainty is statistical and the second is systematic, and the upper
limit is at a $90\%$ confidence level. 
The significance of the $\BR(\Bp\to\Kzb\Kp)$ and $\BR(\Bz\to\KzKzb)$ results are
$3.5\sigma$ and $4.5\sigma$, respectively, including systematic uncertainties.
In addition, we obtain a measurement
of the \CP-violating asymmetry for the $\Bp\to\Kz\pip$ mode and we 
determine a $90\%$ confidence-level interval for the asymmetry in the $\Bp\to\Kzb\Kp$ mode:
${\cal A}_{CP}(\Bp\to\Kz\pip) = -0.087 \pm 0.046 \pm 0.010$ and 
${\cal A}_{CP}(\Bp\to\Kzb\Kp) \in [-0.43, 0.68]$.
\end{abstract}

\date{\today}

\vfill

\bigskip
\begin{center}
  Submitted to the 32$^{\rm nd}$ International Conference on High-Energy Physics, ICHEP 04,\\
  16 August---22 August 2004, Beijing, China
\end{center}

\vspace{1.0cm}
\begin{center}
{\em Stanford Linear Accelerator Center, Stanford University, 
Stanford, CA 94309} \\ \vspace{0.1cm}\hrule\vspace{0.1cm}
Work supported in part by Department of Energy contract DE-AC03-76SF00515.
\end{center}

\newpage
}

\begin{center}
\small

The \babar\ Collaboration,
\bigskip

%
B.~Aubert,
R.~Barate,
D.~Boutigny,
F.~Couderc,
J.-M.~Gaillard,
A.~Hicheur,
Y.~Karyotakis,
J.~P.~Lees,
V.~Tisserand,
A.~Zghiche
\inst{Laboratoire de Physique des Particules, F-74941 Annecy-le-Vieux, France }
% ...

\end{center}\newpage

\section{\boldmath Introduction}
\label{sec:Introduction}
The decays of $B$ mesons into charmless hadronic final states provide important
information for the study of \CP\ violation.  In particular, the study of
the two-body decays $B\to\pi\pi$, $B\to K\pi$, and $B\to KK$ provides 
crucial ingredients for measuring or constraining the values of the 
angles $\alpha$ and $\gamma$, defined as the following ratios of elements of the Cabibbo-Kobayashi-Maskawa quark-mixing matrix~\cite{CKM}:
$\alpha \equiv
{\rm arg}\left[-V^{}_{td}V^*_{tb}/V^{}_{ud}V^*_{ub}\right]$ and
$\gamma \equiv {\rm arg}\left[-V^{}_{ud}V^*_{ub}/V^{}_{cd}V^*_{cb}\right]$.  In this paper, we present measurements of the branching fractions of $B$-meson
decays to the charmless two-body final states $\Kz\pip$, $\Kzb\Kp$,
and $\KzKzb$.~\footnote{Unless explicitly stated otherwise, charge-conjugate 
decay modes are assumed throughout this paper and branching fractions
are averaged accordingly.} For the $\Bp\to\Kz\pip$ and 
$\Bp\to\Kzb\Kp$ modes we also report measurements of the direct \CP
asymmetries in the decay rates,
\begin{equation}
{\cal A}_{C\!P}=\frac{
\Gamma\left(\Bm\to \KS h^-\right)-\Gamma\left(\Bp\to \KS h^+\right)}
{\Gamma\left(\Bm\to \KS h^-\right)+\Gamma\left(\Bp\to \KS h^+\right)}
\,,
\end{equation}
where $h=K,\pi$.

Measurements of the rates and charge asymmetries for $B\to K\pi$ decays can be
used to establish direct \CP\ violation and to constrain the angle 
$\gamma$~\cite{kpigamma}.
The decay $\Bp\to \Kz\pip$ is dominated by the $b\to s$-penguin process and
in the Standard Model (SM) is expected to have ${\cal A}_{CP}<1\%$~\cite{acpkspi}.
Thus, an observation of a sizable 
charge asymmetry could be an indication of non-SM contributions to the
penguin-loop amplitude~\cite{acpkspi,newphys}.
The previously unobserved $B \to K\Kb$ decays proceed via 
penguin and $W$-exchange processes similar to those in 
$\Bz\to\pip\pim$ and can help in the
determination of $\alpha$ in the measurement of
time-dependent CP asymmetries in $\Bz\to\pip\pim$~\cite{kkburas}.
Measurements of the
branching fractions for these decay modes also provide important 
information regarding rescattering processes~\cite{kkrosner}.

\section{\boldmath The \babar\ detector and dataset}
\label{sec:babar}

The measurements presented in this paper are based on data collected
with the \babar\ detector~\cite{bbrnim} at the \pep2\ 
asymmetric-energy $\epem$ collider~\cite{pepii}, located at the 
Stanford Linear Accelerator Center.
The sample consists of $226.6\pm 2.5$ million \BB\ pairs produced at the
\FourS\ resonance (``on-resonance''), which corresponds to an integrated
luminosity of about $205~\invfb$.
An additional $16\invfb$ of data recorded at an \epem\ 
center-of-mass (CM) energy approximately $40\mev$ below the \FourS\ resonance
(``off-resonance'') is used for background studies.

The \babar\ detector is
described in detail in~\cite{bbrnim}.  Charged-particle momenta 
are measured in a tracking system consisting of a five-layer, double-sided
silicon vertex detector and a 40-layer drift chamber (DCH), which operate
in a solenoidal magnetic field of $1.5\,{\rm T}$.  Particles are identified as
pions or kaons based on the Cherenkov angle measured with a detector of
internally reflected Cherenkov light (DIRC).  The direction and energy
of photons are determined from the energy deposits in a segmented
CsI(Tl) electromagnetic calorimeter (EMC).

\section{\boldmath Analysis method}
\label{sec:Analysis}

Hadronic events are selected on the basis of charged-particle multiplicity and 
event topology. We reconstruct $B$-meson candidates decaying to $\Kz X$,
where $X$ can be $\pip$, $\Km$ or $\Kzb$. The $\Kz$ candidates
are reconstructed in the mode $\Kz\to\KS\to\pip\pim$.

The following selection criteria are applied to the candidate $B$-decay products.
Charged tracks are required to be within the tracking fiducial volume 
and to have at least $12$ DCH hits and a minimum transverse momentum of 
$0.1\gevc$. Tracks that are not $\KS$-decay products are
required to originate from the interaction point, to be associated with
at least six Cherenkov photons in the DIRC, and to have a Cherenkov angle
within $4\,\sigma$ of the expected value for a pion or kaon hypothesis.
Candidate \KS\ mesons are reconstructed from pairs of 
oppositely charged tracks that are consistent with originating from a common vertex,
have an invariant mass within $\pm 11.2\mevcc$ of the nominal
$\KS$ mass, and have a measured proper decay time greater than five times its
uncertainty.

The $B$-meson candidate is characterized by two nearly uncorrelated
kinematic variables:  the energy-substituted mass,
$\mes = \sqrt{\left(s/2 + {\mathbf {p}}_i\cdot 
{\mathbf {p}}_B\right)^2/E_i^2- p_B^2}$, and the energy difference,
$\Delta E = E^*_B - \sqrt{s}/2$, where the subscripts~$i$ and $B$ refer to the
initial $\epem$ system and the $B$ candidate, respectively. The asterisk 
denotes the \FourS\ rest frame (CM frame), and $\sqrt{s}$ is the total CM energy.
The pion mass is assigned to all charged particles in the calculation of $E^*_B$.  
For $\Bz\to\Kz\Kzb$ candidates, we require $|\Delta E| < 0.1\gev$, while
for $\Bp\to\Kz h^+$ candidates $(h = \pi,K)$
we require $-0.115 < \Delta E < 0.075\gev$. The interval is asymmetric
in order to select both $\Bp\to\Kz\pip$ and $\Bp\to\Kzb \Kp$ decays
with nearly $100\%$ efficiency. The $\Delta E$ 
distribution peaks near zero in the modes not containing charged kaons. 
In $\Bp\to\Kzb \Kp$ decays, the $\Delta E$ distribution peaks at $-45\mev$
as a result of the assignment of the pion mass to the charged kaon candidate.
The distribution of $\mes$ peaks at the $B$ mass in all modes.  We impose a loose selection of $5.20 < \mes < 5.29\gevcc$ that includes a background-dominated region used to estimate the level of backgrounds in the signal region.

Simulated events~\cite{bbrgeant}, off-resonance data, and events in 
on-resonance $\mes$- and $\Delta E$-sideband regions are used to study
backgrounds. The contribution from other $B$-meson decays is found to be
negligible.
The primary source of background are random combinations of tracks and neutral
clusters produced in $\epem \to \qqbar$ events, where $q$ is a $u$, $d$,
$s$, or $c$ quark. In the CM frame, these ``combinatorial'' events are
characterized by a jet-like structure, in contrast to the more uniformly
distributed decays of $B$ mesons produced in \FourS\ decays.
We suppress combinatorial backgrounds by exploiting this topological difference
through a selection based on $\theta_S^*$, the angle between the sphericity axis of the $B$-candidate decay products and the sphericity axis of the remaining particles in the event. 
In the CM frame, this selection is $|\cos\theta_S^*| < 0.8$.
To discriminate further between signal decays and combinatorial backgrounds,
we employ a Fisher discriminant, $\cal F$.
We define $\cal F$ as an optimized linear combination of $\sum_i p^*_i$ and
$\sum_i p^*_i \cos^2 \theta^*_i$~\cite{alphaprl}, where $p^*_i$ is the
momentum of particle $i$ and $\theta^*_i$ is the angle between its momentum
and the thrust axis of the $B$-candidate decay products, both calculated in the CM frame.
The sums are over all particles in the event except for the $B$-candidate
decay products. The difference between signal and background distributions
of $\cal F$ is present in the probability density functions (PDF's) that model
$\cal F$ in the fits to the data sample, which we describe below.

Signal yields and charge asymmetries are determined from unbinned extended
maximum-likelihood fits.  The extended likelihood for a sample of
$N$ $\Kz X$ candidates is defined as
\begin{equation}
{\cal L} = \exp{\left(-\sum_{i}n_i \right)}
\prod_{j=1}^N\left[\sum_i N_i{\cal P}_i(\vec{x}_j;\vec{\alpha}_i)\right],
\end{equation}
where ${\cal P}_i(\vec{x}_j;\vec{\alpha}_i)$ is the PDF of a
signal or background category $i$, evaluated at the values of the measured variables
$\vec{x}_j$ of candidate $j$. The sum is over the set of categories.
The parameters $\vec{\alpha}_i$ determine the expected distributions of 
measured variables in each category, and $n_i$ are the yields being determined in
the fit. The probability coefficients $N_i$ are defined separately for each mode.
We perform separate fits to the two samples of $B$ candidates:
$\Bz \to\Kz\Kzb$ and $\Bp \to\Kz h^+$ ($h=\pi$, $K$).
In the fit to the ``neutral-$B$'' ($\Bz \to\Kz\Kzb$) sample we include two categories,
signal and background.  The probability coefficients $N_i$ are set equal to the yields
($N_i=n_i$); the yield in each category is obtained by maximizing the likelihood.
In the fit to the ``charged-$B$'' ($\Bp\to \Kz h^+$) sample, we fit simultaneously
two signal categories, $\Bp \to \Kz \pip$ and $\Bp \to \Kzb \Kp$, and two
corresponding background categories.  In addition, the probability
coefficient for each category $i$ is given by 
$N_i = n_i\left(1-q_j{\cal A}_i\right)$, where
$n_i$ is the $\rm total$ yield, summed over charge states, ${\cal A}_i$ is 
the charge asymmetry, and $q_j$ is the charge of the $B$
candidate.  The total yields and charge asymmetries are determined by 
maximizing ${\cal L}$.

As the input variables in the fit are nearly uncorrelated, the PDF
in the likelihood function, ${\cal P}_i(\vec{x}_j;\vec{\alpha}_i)$,
is constructed as the product of the individual PDF's of the input
variables $\vec{x}_j$.  In both fits, the set of input variables contains
\mes, $\Delta E$, and $\cal F$.
In the fit to the $B^+\to \Kz h^+$ sample, we include also the normalized Cherenkov-angle residuals 
$\left( \theta_c - \theta_c^\pi\right)/\sigma_{\theta_c}$ and
$\left( \theta_c - \theta_c^K\right)/\sigma_{\theta_c}$, where $\theta_c$ is
the measured Cherenkov angle of the primary daughter $h^+$, 
$\sigma_{\theta_c}$ is its measurement uncertainty, and
$\theta_c^\pi\,(\theta_c^K)$ is the expected Cherenkov angle for a pion (kaon) hypothesis.
The quantities $\sigma_{\theta_c}$, $\theta_c^\pi$, and $\theta_c^K$ are 
measured separately for negatively and positively charged pions and kaons
from a control sample of $\Dz\to \Km\pip$ decays originating from $\Dstarp$ decays.

The parameterization of the PDF's is determined from
data and Monte Carlo-simulated (MC) samples. Some PDF parameters are free
to vary in the fit as explained below.
The signal \mes\ PDF's for $\Bp\to \Kz h^+$ and $\Bz \to \Kz\Kzb$
are derived from signal MC samples.
The shape of the distribution is modeled as a Gaussian function
with an asymmetric variance and a low-side power tail
in the $\Kz h^+$ fit, while in the $\Kz\Kzb$ fit it is parameterized as
a linear combination of two Gaussian functions
(``double-Gaussian'' function).
The mean value of the signal \mes\ distribution
is a free parameter in the $\Bp\to \Kz h^+$ fit,
as the sample of candidates is sufficiently large.
To describe the background \mes\ PDF,
we use an empirical threshold function~\cite{argusfcn}.
The shape parameter of this function
is a free parameter in the $\Bp\to \Kz h^+$ fit, while
in the $\Bz \to \Kz\Kzb$ fit it is determined from the on-resonance $\Delta E$-sideband region,
defined by the selection $0.1 < |\Delta E| < 0.3\gev$.

The signal $\Delta E$ PDF's are derived from MC samples and
modeled as double-Gaussian functions for both modes.
The signal $\Delta E$ distribution is expected to be centered
near zero for the $\Kz \pip$ mode, while for $\Kzb \Kp$ candidates,
the mean of $\Delta E$ is shifted because the pion mass is assumed
for the charged track.
The shift in $\Delta E$ is
\begin{equation}
\nonumber
\langle \Delta E \rangle = -\gamma_{\rm boost}\times
\left( \sqrt{M_K^2+p^2}-\sqrt{M_\pi^2+p^2} \right)\, ,
\end{equation}
where $p$ is the momentum of the track, and $M_\pi$ and $M_K$
are the nominal values of the pion mass and the kaon mass, respectively.
The $\Delta E$ mean value is a free parameter in the
$\Bp\to \Kz h^+$ fit, while in the $\Bz \to \Kz\Kzb$ fit
it is determined through a
comparison of the values obtained in the two MC samples
with the value obtained in the fit to the
$\Bp\to \Kz h^+$ data sample.
The background $\Delta E$ distribution is modeled as a
second- and first-degree
polynomial function for the charged-$B$ and neutral-$B$ modes,
respectively. The polynomial coefficients are determined
from on-resonance events in the $\mes$-sideband region,
defined by the selection $5.20 < \mes < 5.26\gevcc$.

In both modes, the signal $\cal F$ distribution
is modeled as a Gaussian function with an asymmetric variance~\cite{asymgauss}.
Its parameters are free to vary in the $\Kz h^+$ fit; in the $\Kz\Kzb$ fit, 
they are determined from the MC sample.
The background $\cal F$ distribution is parameterized as a
double-Gaussian function with its parameters left free to vary in both fits.

In the charged-$B$ modes, the normalized Cherenkov-angle residuals
are modeled as double-Gaussian functions; the PDF parameters are
taken from the $D^*$ control sample and 
they are determined separately for $\pip$, $\pim$, $\Kp$, and $\Km$
tracks as a function of momentum and polar angle.

\renewcommand{\multirowsetup}{\centering}
\newlength{\LL}\settowidth{\LL}{$20441$}
\begin{table}[!tb]
\caption{\small Summary of results for numbers of selected $\Kz X$ 
candidates $N$, total 
detection efficiencies $\eps$, fitted signal yields $N_S$, statistical
significances $S$ including systematic uncertainties, charge-averaged
branching fractions $\BR$, and charge asymmetries ${\cal A}_{CP}$. The
efficiencies include the branching fractions of intermediate states
($\Kz\to\KS\to\pip\pim$).
Branching fractions are calculated assuming equal rates 
for the $\upsbzbz$ and $\FourS \to \Bp\Bm$ processes.
The upper limit for the $\Kzb\Kp$ branching
fraction corresponds to a $90\%$ confidence-level (C.L.), and the central value is given
in parentheses.
The $90\%$ C.L. asymmetry interval is given for
$\Kzb\Kp$, including the systematic uncertainty.
The $90\%$ C.L. asymmetry interval for $\Kz\pip$ is
$[-0.164, -0.010]$.
}
\label{tab:results}
\begin{center}
\hspace*{-0.4cm}
{\footnotesize
\begin{tabular}{lccccccc}
\hline\hline
Mode  & $N$ & $\eps$ (\%) & $N_{S}$ & $S(\sigma)$ & \BR($10^{-6}$) & 
${\cal A}_{CP}$  \\
\hline \\[-3mm]
$\Kz\pip$  & \multirow{2}{\LL}{$20441$} & $12.6 \pm 0.3$ & $744$ $^{+37}_{-36}$ $^{+21}_{-17}$  &
            $-$  & $26.0 \pm 1.3 \pm 1.0$ & $-0.087 \pm 0.046 \pm 0.010$\\[2mm]
$\Kzb\Kp$  &                           & $12.5 \pm 0.3$ & $41$ $^{+15}_{-13}$ $^{+3}_{-2}$ &
            $3.5$ & $<2.35 \, (1.45 ^{+0.53}_{-0.46} \pm 0.11)$ & $[-0.43, 0.68]$ \\[2mm]
$\KzKzb$   & $1939$                     & $8.5 \pm 0.5$ & $23.0$ $^{+7.7}_{-6.7}$ $^{+1.9}_{-2.0}$ &
            $4.5$ & $1.19^{+0.40}_{-0.35}\pm 0.13$ &$-$& \\
\hline\hline
\end{tabular}
}
\end{center}
\end{table}

\begin{figure}[!tbp]
\begin{center}
\includegraphics[width=0.45\linewidth]{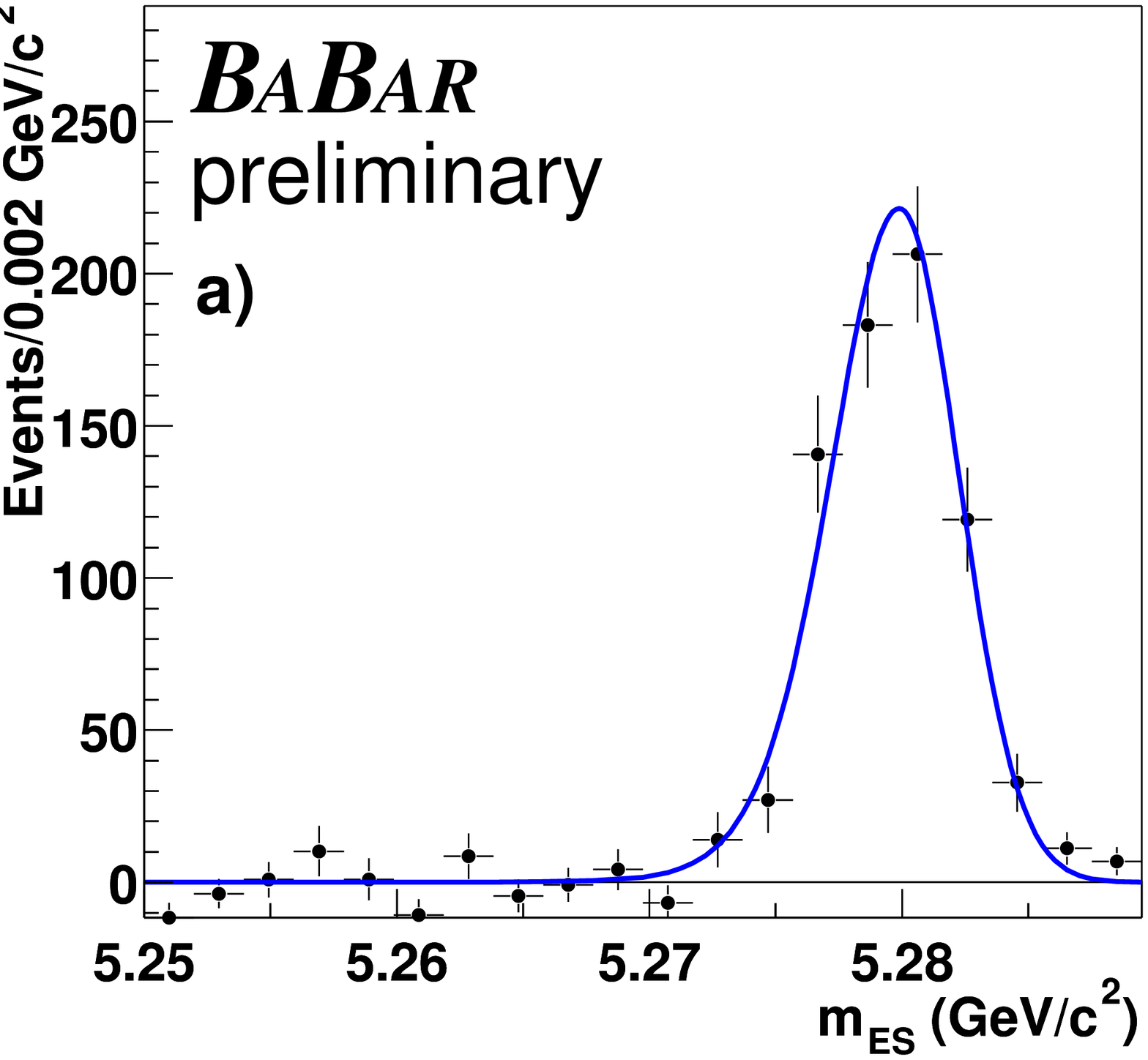}
\includegraphics[width=0.45\linewidth]{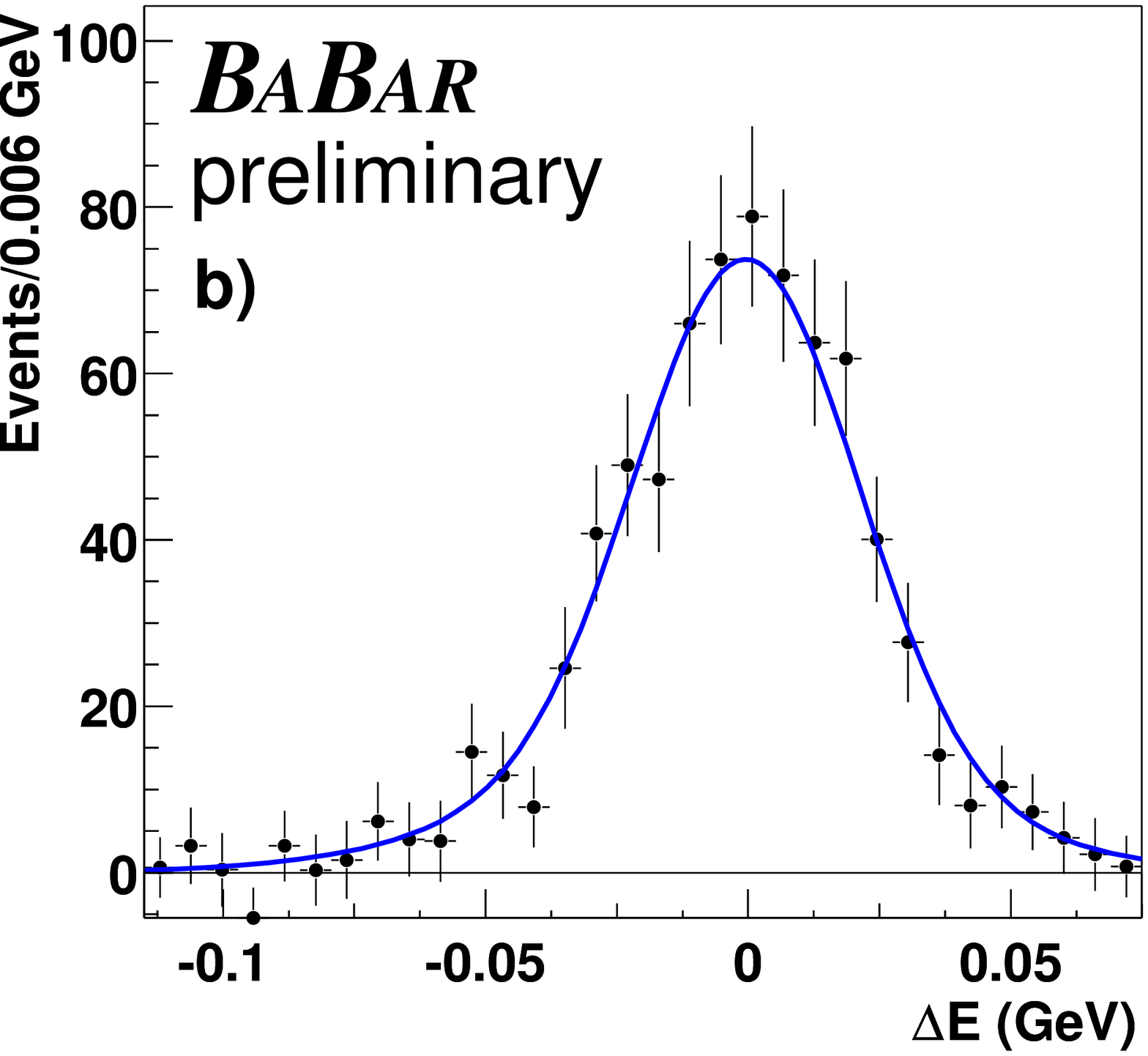}
\includegraphics[width=0.45\linewidth]{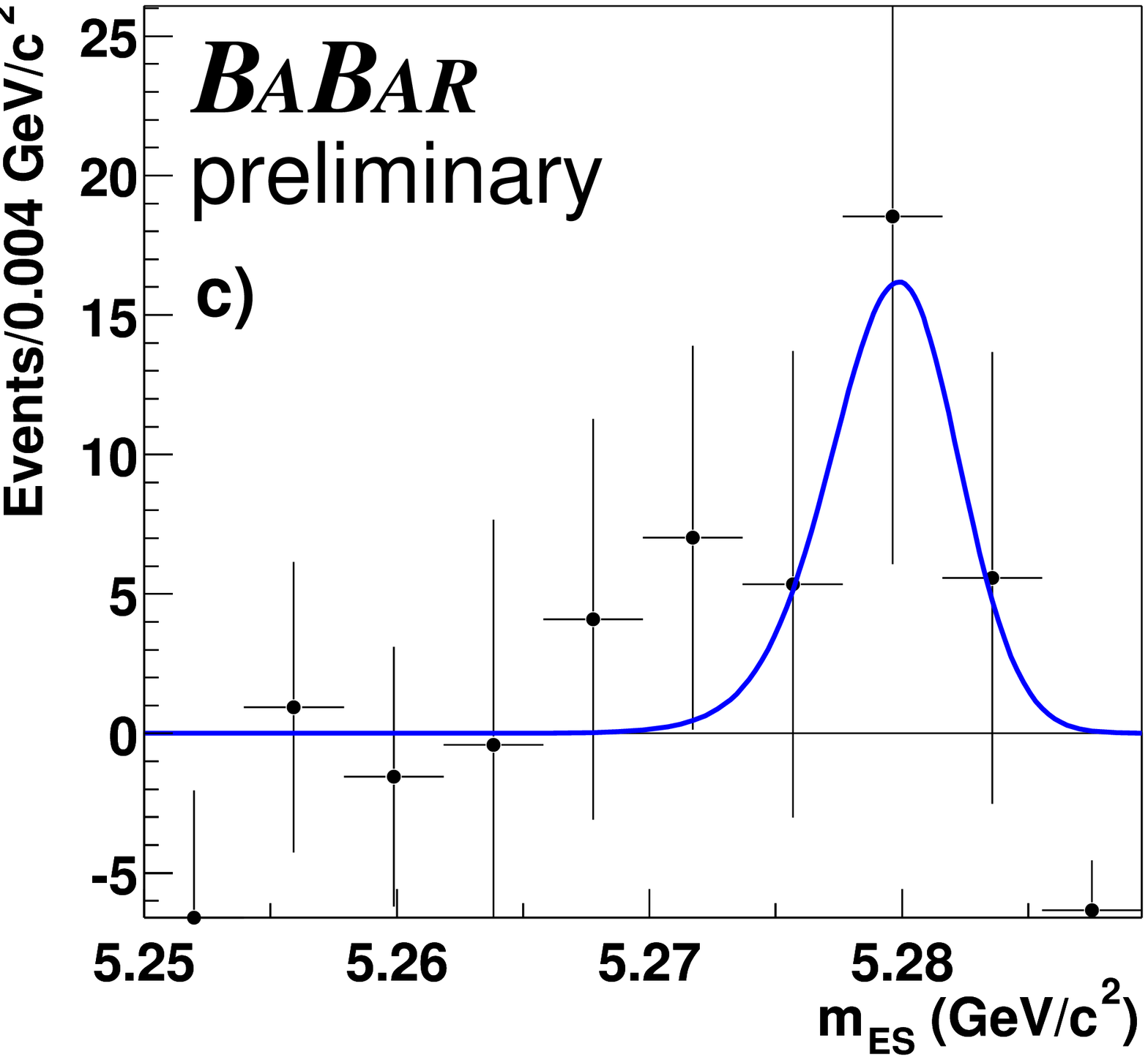}
\includegraphics[width=0.45\linewidth]{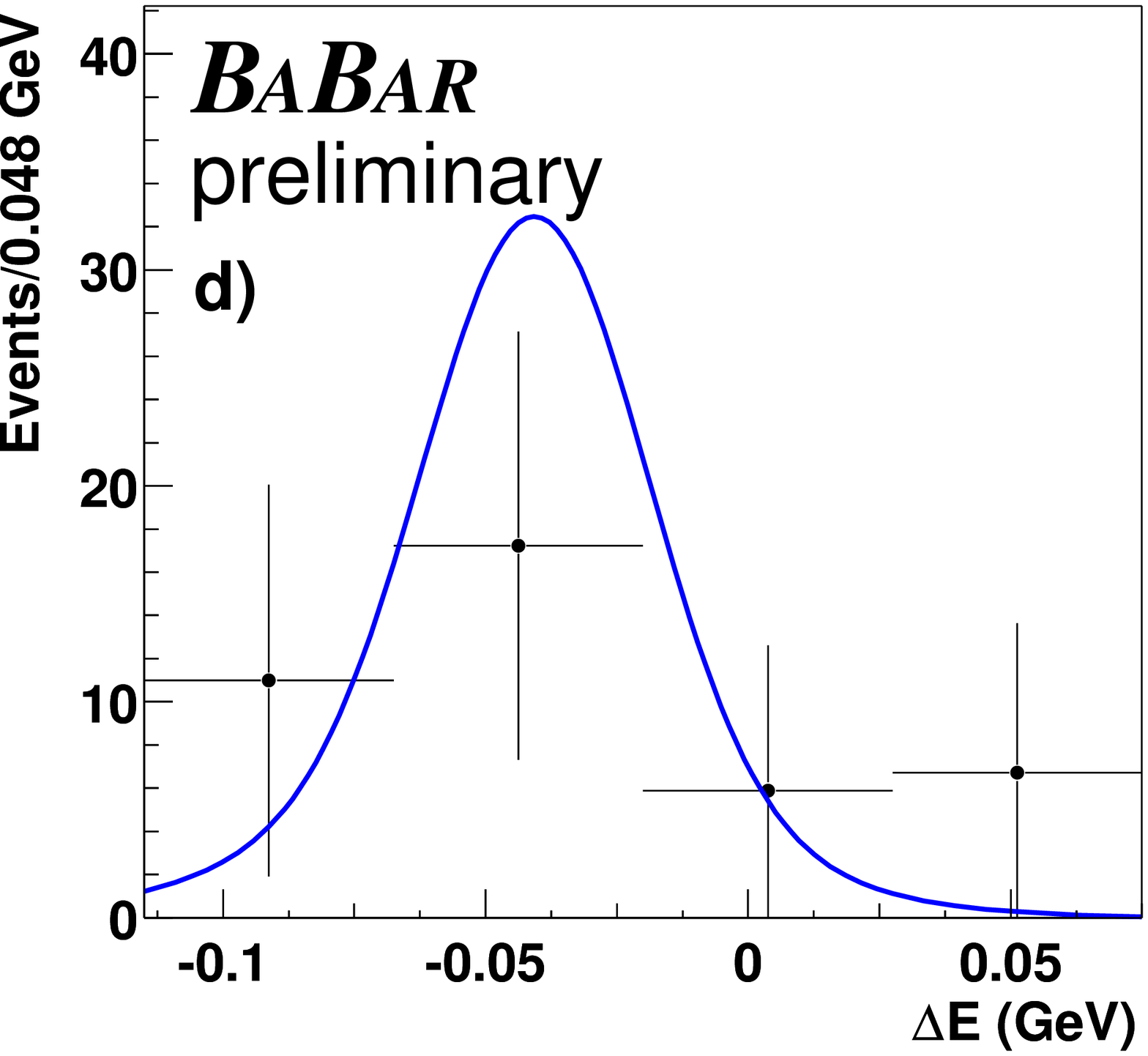}
\includegraphics[width=0.45\linewidth]{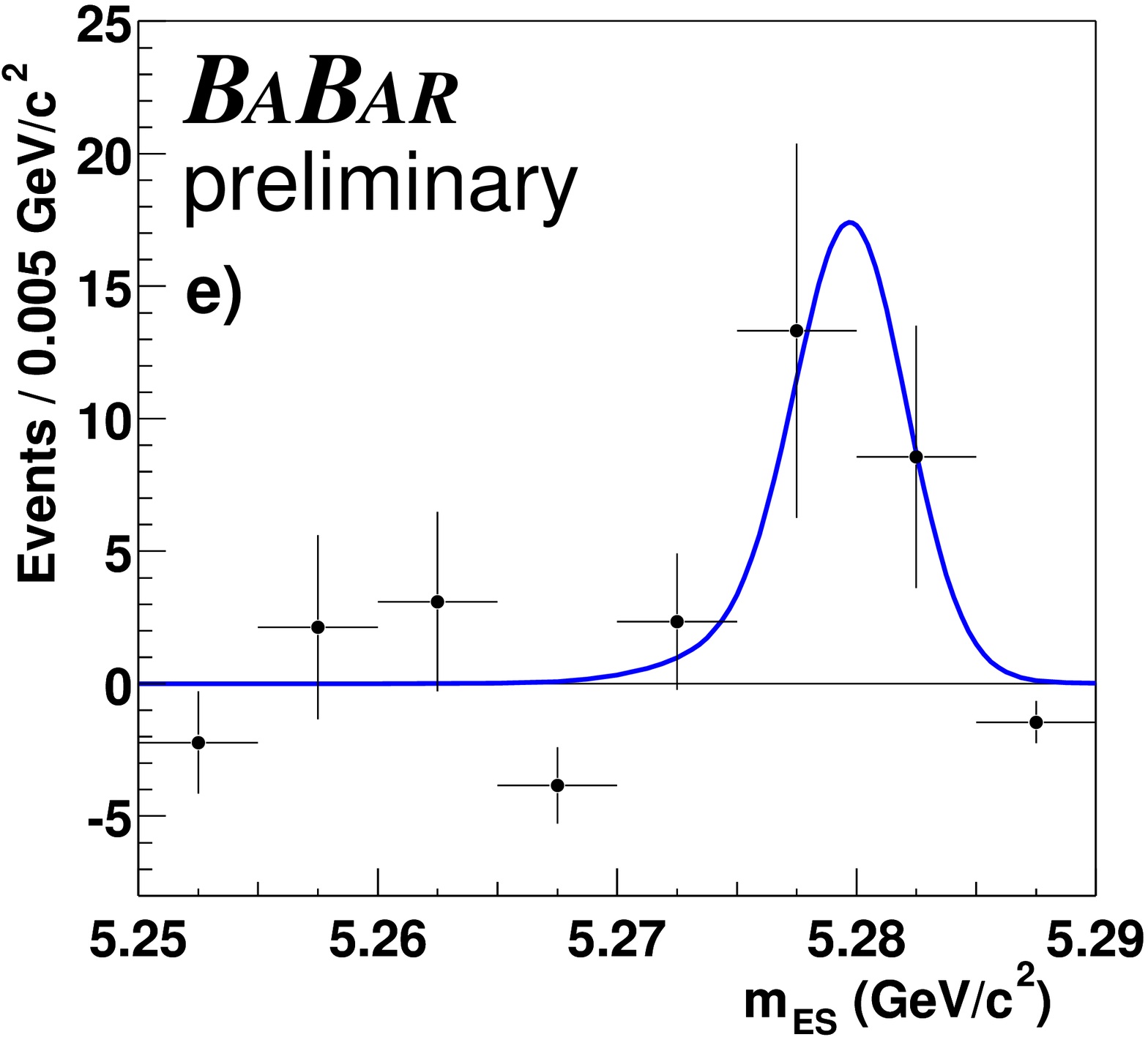}
\includegraphics[width=0.45\linewidth]{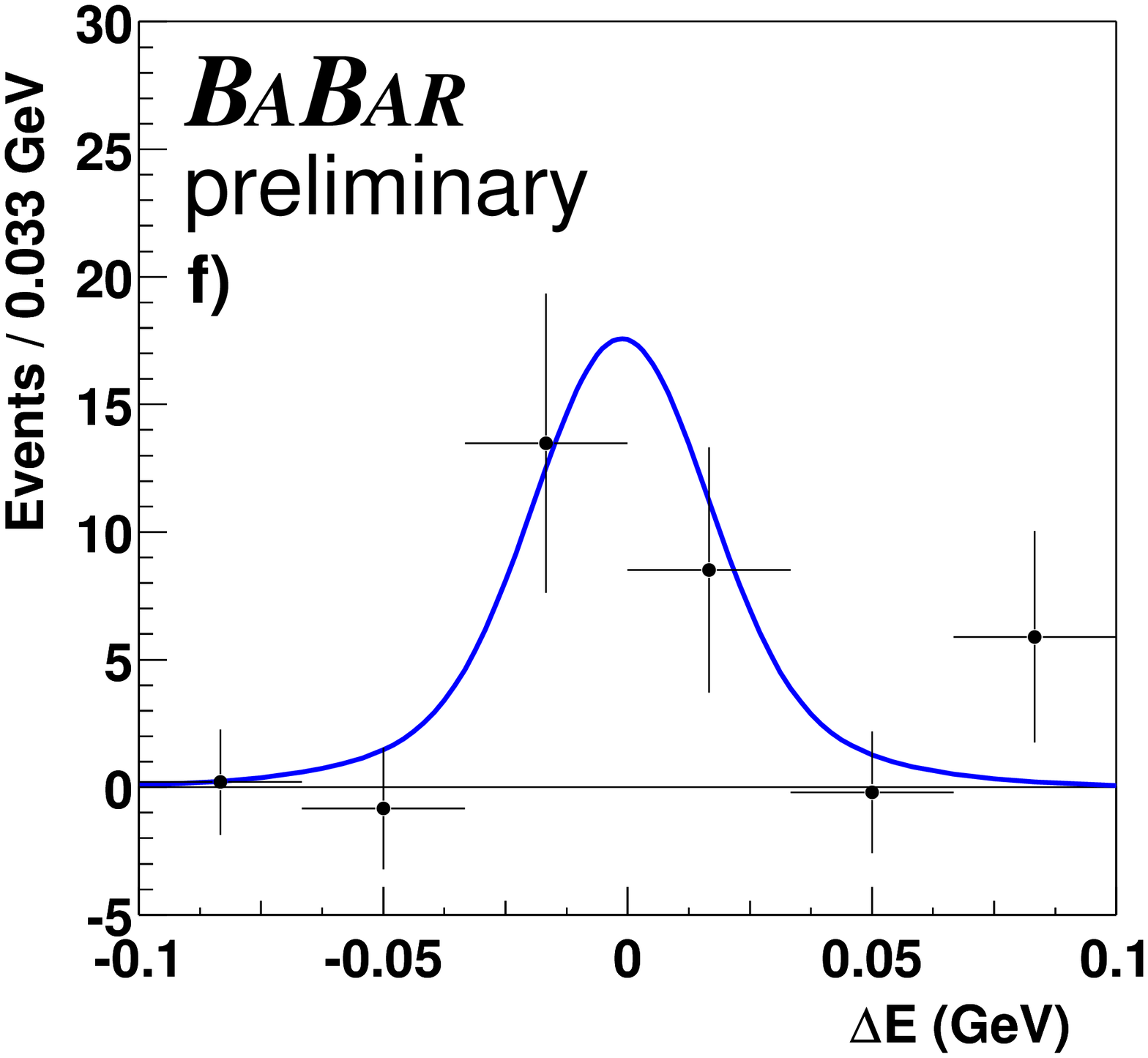}
\caption{\small \small Distributions of $\mes$ and $\Delta E$ for
(a,b) $\Bp\to \KS\pip$, (c,d) $\Bp\to\KS\Kp$ and (e,f) $\Bz\to\KS\KS$
(histograms) after background subtraction (see text). 
Projections of the fit PDF's are overlaid (solid curves).}
\label{fig:mesde}
\end{center}
\end{figure}

\begin{figure}[!tbp]
\begin{center}
\includegraphics[width=0.45\linewidth]{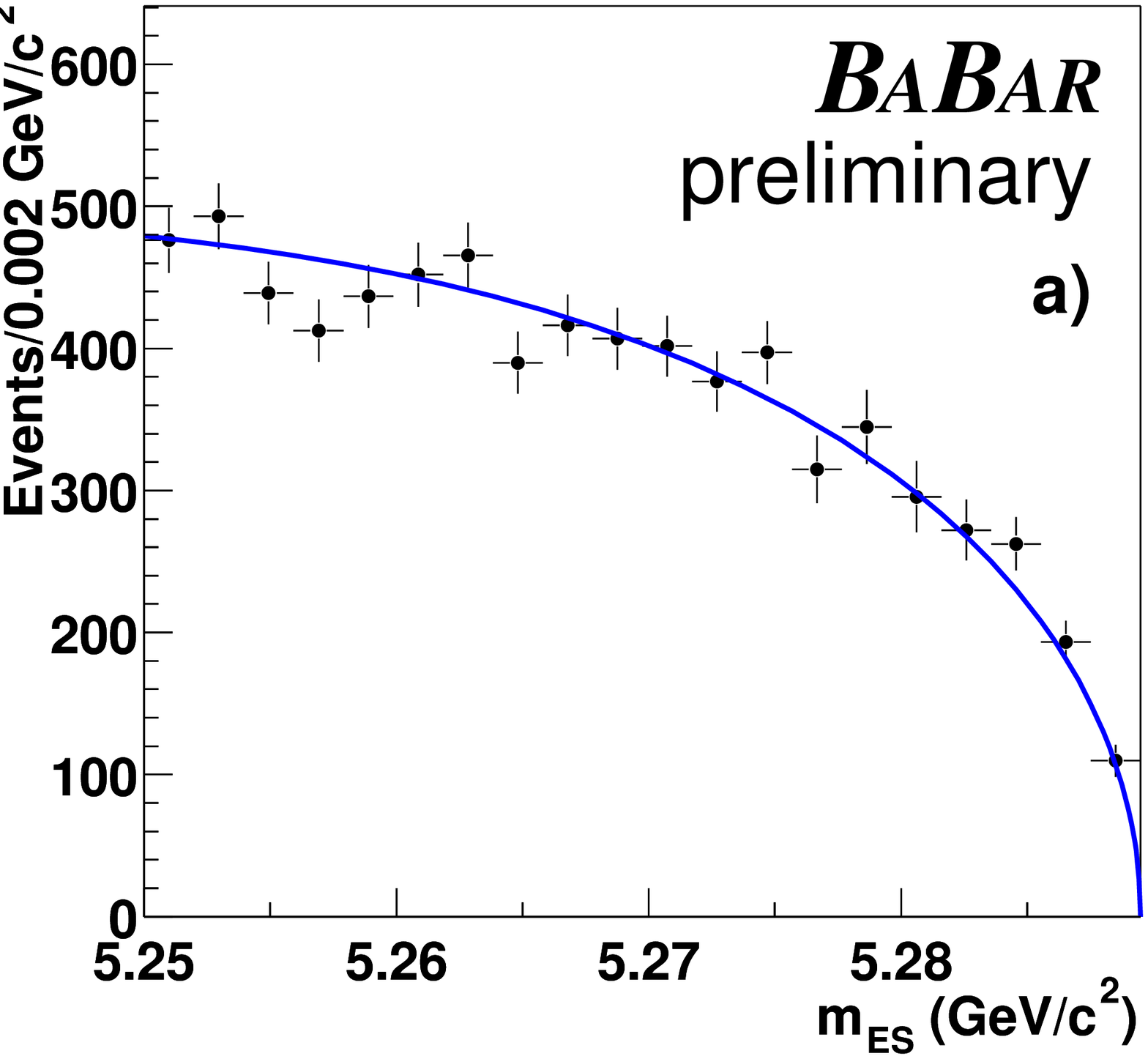}
\includegraphics[width=0.45\linewidth]{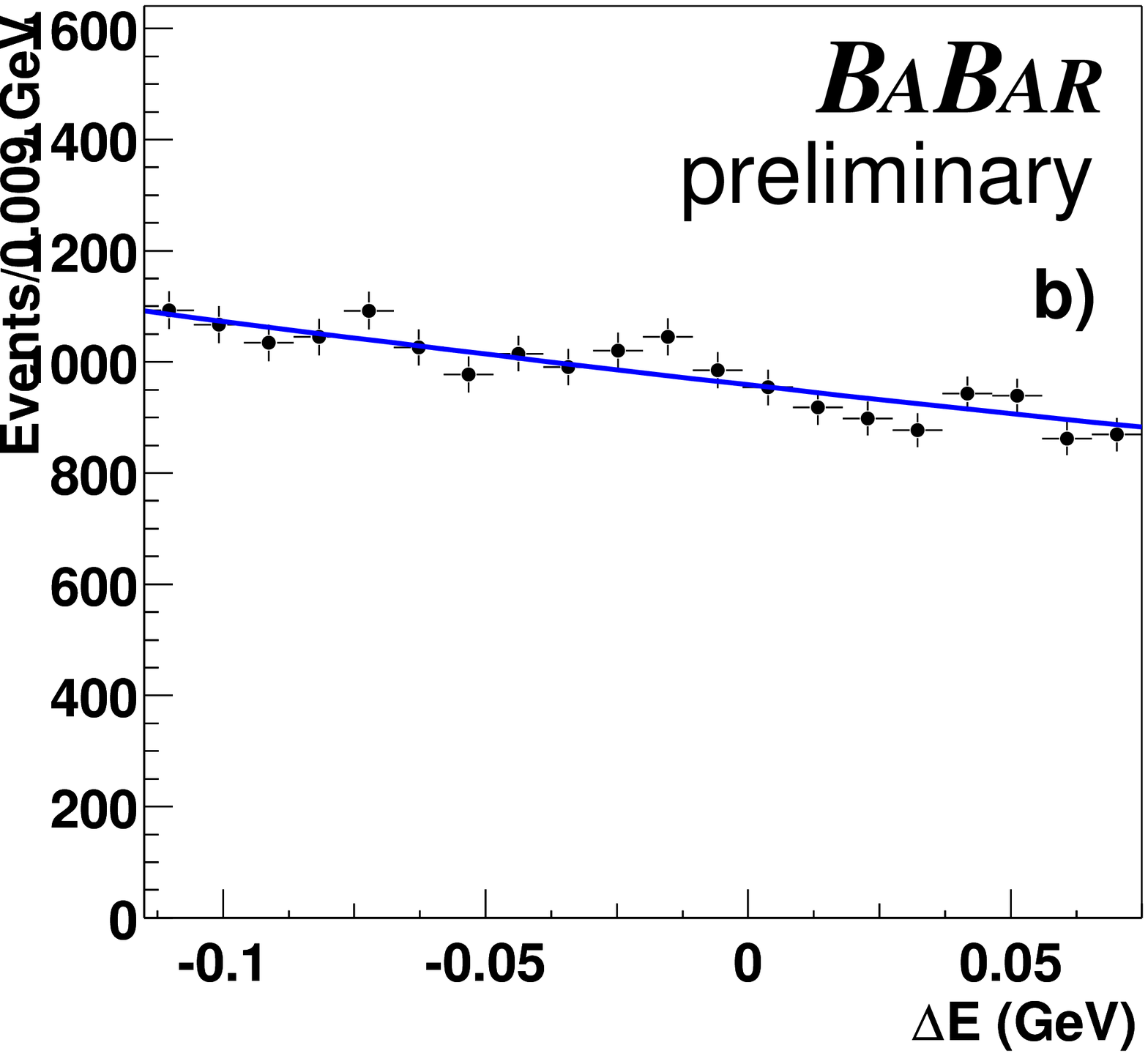}
\includegraphics[width=0.45\linewidth]{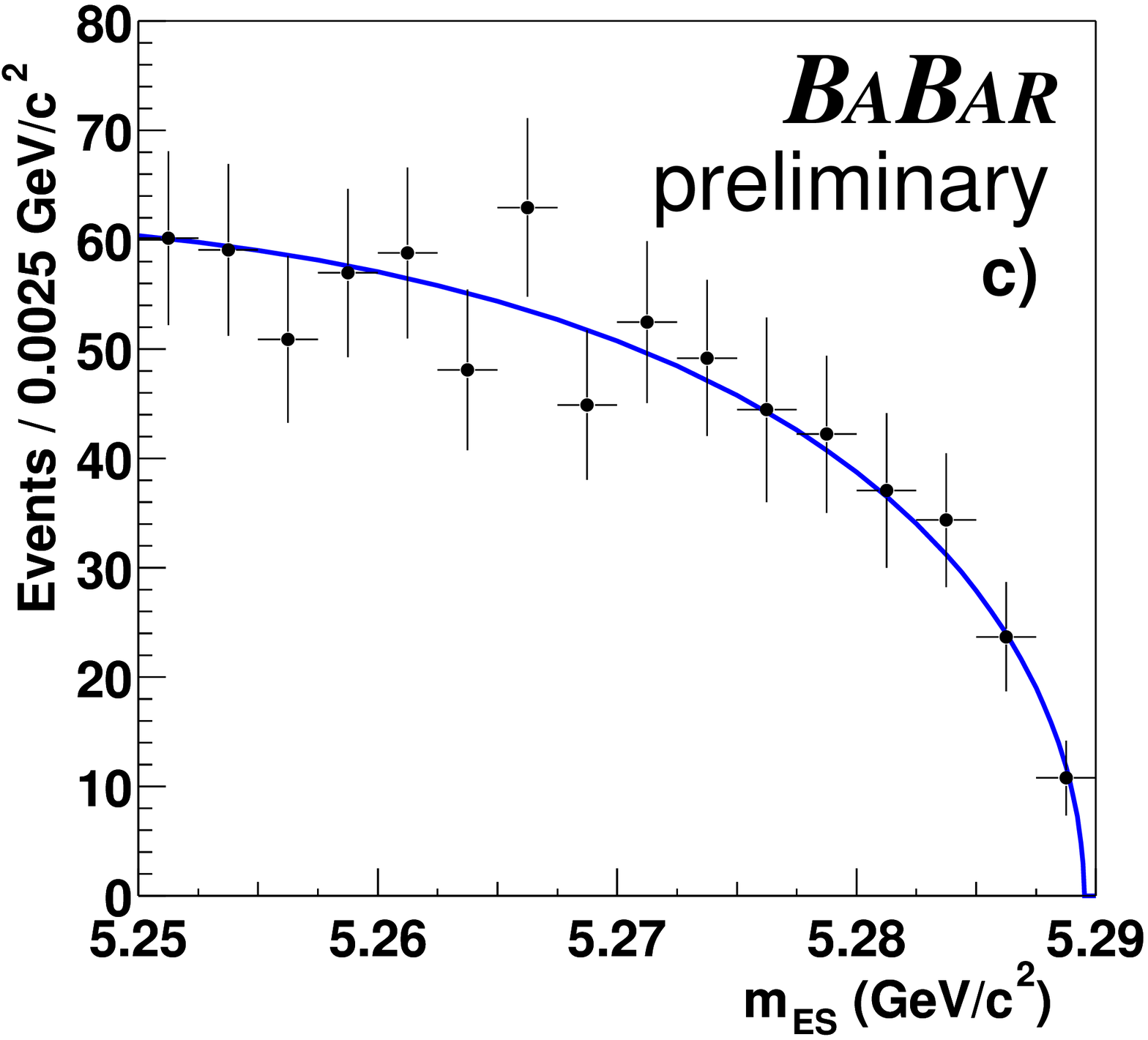}
\includegraphics[width=0.45\linewidth]{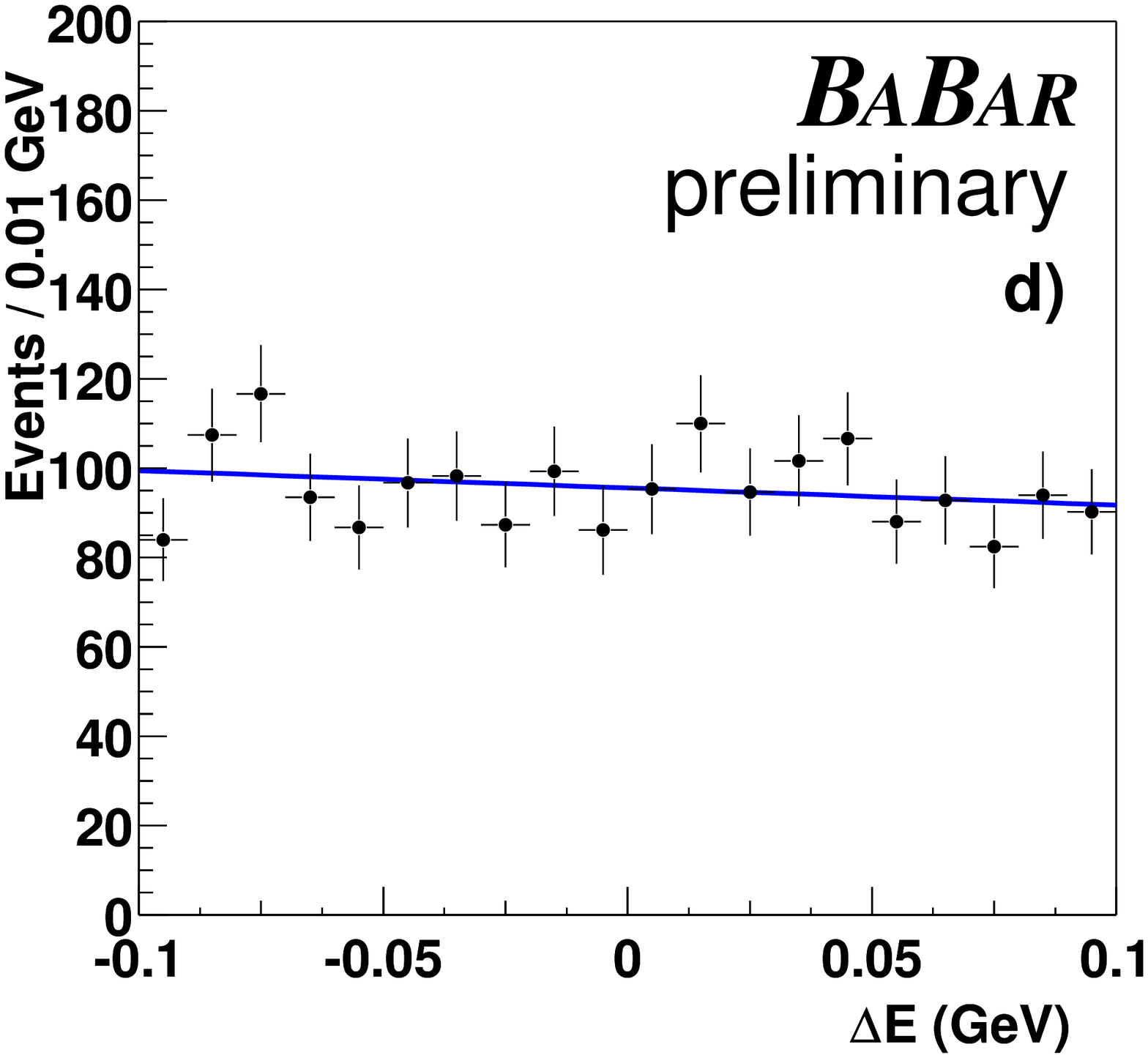}
\caption{\small \small Distributions of $\mes$ and $\Delta E$ for
(a,b) $\Bp\to \KS h^{+}$ and (c,d) $\Bz\to\KS\KS$
(histograms) after signal subtraction (see text). 
Projections of the fit PDF's are overlaid (solid curves).}
\label{fig:mesde-bkg}
\end{center}
\end{figure}

\section{\boldmath Physics results and systematic uncertainties}
\label{sec:Physics}

The results of the two maximum-likelihood fits are summarized in Table~\ref{tab:results}.
The $\KzKzb$ final state is an equal admixture
of $\KS\KS$ and $\KL\KL$. We therefore use a $50\%$ probability
for the $\KzKzb$ to decay as $\KS\KS$ in computing the $\Bz\to\KzKzb$
branching fraction. We also use the current world averages for
${\cal B}(\KS\to\pip\pim)$~\cite{PDG2004}.

Figure~\ref{fig:mesde} shows background-subtracted distributions of $\mes$ and
$\Delta E$ for $\Bp\to \Kz h^{+}$ and $\Bz\to\Kz\Kzb$ candidates.
The background subtraction is performed by weighting events using the
$_{s}\cal P$$lot$ technique described in Ref.~\cite{sPlot}.
The shape of the resulting distribution can be compared with the PDF used in the full fit.
We find good agreement in both variables for $\Kz h^{+}$ and $\Kz\Kzb$ candidates.
The corresponding signal-subtracted distributions of $\mes$ and $\Delta E$
are shown in Figure~\ref{fig:mesde-bkg}.

The signal significance is defined as the square root of 
the difference in $-2\ln{\cal L}$ between the best fit and the 
null-signal hypothesis. The upper limit on the signal yield for a given mode $i$ is defined as the value
of $n_i^{\rm UL}$ for which 
$\int_0^{n_i^{\rm UL}}{\cal L}_{\rm max}dn_i/
\int_0^\infty {\cal L}_{\rm max}dn_i = 0.9$, where
$\cal L_{\rm max}$ is the likelihood as a function of $n_i$, maximized with 
respect to the remaining fit parameters. Branching-fraction upper limits are
then calculated by increasing the signal-yield upper limit and reducing the
efficiency by their respective systematic uncertainties.

Systematic uncertainties in the signal yields arise primarily from imperfect
knowledge of the PDF parameterizations. Such systematic errors are evaluated either by
varying the PDF parameters by their measured ($1\sigma$) uncertainties
or by substituting alternate parameterizations.  The significance of each signal yield with the systematic uncertainties included is evaluated by imposing simultaneously in the fit all of the
systematic variations of the PDF's that lower that signal yield.
Systematic uncertainties in the efficiency include uncertainties in
tracking and $\KS$ reconstruction.

In the $\Kz h^+$ sample, the dominant systematic uncertainty is
that associated with the signal $\mes$ shape, leading to a systematic error of $^{+13}_{-15}$ ($^{+1.3}_{-1.7}$)
events, and that associated with the signal $\Delta E$ resolution, leading to a systematic error of $^{+16}_{-5}$ ($^{+2.8}_{-0.7}$)
events in the $\Bp\to \Kz\pip$ ($\Kzb\Kp$) mode.
The significance of the $\Bp\to \Kz\Kp$ observation with (without) the systematic
uncertainty included is $3.5\,\sigma$ ($3.7\,\sigma$).
In the $\Bz\to \Kz\Kzb$ mode, the main systematic uncertainty is that
associated with the background $\mes$ distribution ($\pm 0.8$ events).
There is also a sizable systematic contribution from the
imperfect agreement between MC and data samples ($\pm 1.4$ events) and,
in the branching-fraction extraction, the uncertainty in the $\KS$ efficiency ($6.0\%$).
The significance of the $\Bz\to \Kz\Kzb$ observation with (without) the systematic
uncertainty included is $4.5\,\sigma$ ($4.8\,\sigma$).
The systematic uncertainties in the charge asymmetries in the $\Bp\to \Kz\pip$
mode are evaluated by adding in quadrature the contributions
from PDF variations and the upper limit on intrinsic charge bias in the
detector ($\pm 0.010$).

\section{Summary}
\label{sec:Summary}

We present preliminary measurements of the branching fraction
and the \CP-violating charge asymmetry in the $\Bp\to\Kz\pip$ decay,
and a preliminary measurement of the branching fraction of the $\Bz\to\KzKzb$ decay.
No evidence of direct \CP\ violation in the $\Bp\to\Kz\pip$ mode is observed.
The $\Bz\to\KzKzb$ measurement constitutes the first observation of this decay channel:
the probability of obtaining our result assuming the null-signal hypothesis is $3.4 \times 10^{-6}$.
We have also searched for the $\Bp\to \Kzb\Kp$ decay and
set an upper limit on its branching fraction at 
$2.35\times 10^{-6}$ at the $90\%$ C.L.
The branching-fraction measurements reported here are consistent with
previous measurements of the same quantities~\cite{cleoprla,hhprl,belleprd,belleacp},
but are extracted from a data sample larger by a factor of $2.6$.

\section{Acknowledgments}
\label{sec:Acknowledgments}

\par
We are grateful for the excellent luminosity and machine conditions
provided by our \pep2\ colleagues, 
and for the substantial dedicated effort from
the computing organizations that support \babar.
The collaborating institutions wish to thank 
SLAC for its support and kind hospitality. 
This work is supported by
DOE
and NSF (USA),
NSERC (Canada),
IHEP (China),
CEA and
CNRS-IN2P3
(France),
BMBF and DFG
(Germany),
INFN (Italy),
FOM (The Netherlands),
NFR (Norway),
MIST (Russia), and
PPARC (United Kingdom). 
Individuals have received support from CONACyT (Mexico), A.~P.~Sloan Foundation, 
Research Corporation,
and Alexander von Humboldt Foundation.

\end{document}